\documentclass[journal=langd5,manuscript=article]{achemso}

\usepackage{graphics}
\usepackage[american]{babel}
\usepackage{epsfig}
\usepackage{amsmath}
\usepackage{amssymb}
\usepackage{array}
\usepackage{slashbox}
\usepackage[version=3]{mhchem}

\newcommand{\bondlength}{\ensuremath{A}}
\newcommand{\bondlengthCC}{\ensuremath{\bondlength}}
\newcommand{\contactangle}{\ensuremath{\vartheta}}

\newcommand{\cpenergy}{\ensuremath{Z}}
\newcommand{\cutoff}{\ensuremath{\distance_\mathrm{c}}}
\newcommand{\degree}[1]{\ensuremath{{#1}^{\circ}}}
\newcommand{\density}{\ensuremath{\rho}}
\newcommand{\densityliq}{\ensuremath{\density'}}
\newcommand{\densitytr}{\ensuremath{\density_\mathrm{vl}}}
\newcommand{\densityvap}{\ensuremath{\density''}}
\newcommand{\differential}{\ensuremath{\mathnormal{d}}}
\newcommand{\distance}{\ensuremath{\mathnormal{r}}}
\newcommand{\FWenergy}{\ensuremath{\mathnormal{\zeta}}}

\newcommand{\inftemperature}{\ensuremath{\mathnormal{\tau}}}
\newcommand{\interlayer}{\ensuremath{\mathnormal{Y}}}
\newcommand{\kboltz}{\ensuremath{\mathnormal{k}}}
\newcommand{\layer}{\ensuremath{\mathnormal{\ell}}}
\newcommand{\layers}{\ensuremath{\mathnormal{L}}}
\newcommand{\LJenergy}{\ensuremath{\varepsilon}}
\newcommand{\LJenergyFW}{\ensuremath{\LJenergy_\mathrm{fw}}}
\newcommand{\LJsize}{\ensuremath{\sigma}}
\newcommand{\LJsizeFW}{\ensuremath{\LJsize_\mathrm{fw}}}
\newcommand{\LJTS}{LJTS}
\newcommand{\mardyn}{\textit{$\ell$s1 Mardyn}}
\newcommand{\moli}{\ensuremath{\mathnormal{i}}}
\newcommand{\molj}{\ensuremath{\mathnormal{j}}}
\newcommand{\molmass}{\ensuremath{\mathnormal{m}}}
\newcommand{\mult}{\ensuremath{\cdot}}
\newcommand{\potential}{\ensuremath{\mathnormal{u}}}
\newcommand{\potentialfw}{\ensuremath{\mathnormal{u}_\mathrm{fw}}}
\newcommand{\potentialCC}{\ensuremath{\potential^\mathnormal{xz}}}
\newcommand{\potentialLJ}{\ensuremath{\potential^\mathrm{LJ}}}
\newcommand{\potentialLJTS}{\ensuremath{\potential^\mathrm{ts}}}
\newcommand{\potentialy}{\ensuremath{\potential^\ycoord}}

\newcommand{\soldensity}{\ensuremath{\mathnormal{\density_\mathrm{w}}}}
\newcommand{\springconstant}{\ensuremath{\mathnormal{D}}}
\newcommand{\springconstantCC}{\ensuremath{\springconstant_\mathnormal{xz}}}
\newcommand{\springconstanty}{\ensuremath{\springconstant_\ycoord}}
\newcommand{\sptwo}{sp$^2$}
\newcommand{\surfacedensity}{\ensuremath{\mathnormal{\eta}}}
\newcommand{\surfacerad}{\ensuremath{\mathnormal{\lambda}}}
\newcommand{\temperature}{\ensuremath{\mathnormal{T}}}
\newcommand{\temperaturecrit}{\ensuremath{\temperature_\mathrm{c}}}
\newcommand{\temperaturew}{\ensuremath{\temperature_\mathrm{w}}}
\newcommand{\tension}{\ensuremath{\gamma}}
\newcommand{\tensionvl}{\ensuremath{\tension}}
\newcommand{\tensions}{\ensuremath{\tension_\mathrm{s}}}
\newcommand{\vapour}{vapor}
\newcommand{\Vapour}{Vapor}
\newcommand{\vapor}{\vapour}

\newcommand{\welldepth}{\ensuremath{\mathnormal{W}}}
\newcommand{\xcoord}{\ensuremath{\mathnormal{x}}}
\newcommand{\ycoord}{\ensuremath{\mathnormal{y}}}
\newcommand{\ycoordadd}{\ensuremath{\mathnormal{\upsilon}}}
\newcommand{\ycoordeq}{\ensuremath{\ycoord^\circ}}
\newcommand{\zcoord}{\ensuremath{\mathnormal{z}}}

\title[Contact angle dependence on dispersion]
{Contact angle dependence on the fluid-wall dispersive energy}

\author{Martin Horsch}
\affiliation[University of Paderborn]
{Thermodynamics and Energy Technology Laboratory (ThEt), University of Paderborn, Warburger Str.\ 100, 33098 Paderborn, Germany}
\author{Martina Heitzig}
\altaffiliation{Current address: Computer Aided Process-Product Engineering Center (CAPEC), Technical University of Denmark, S\o{}ltofts Plads, 2800 Kgs.\ Lyngby, Denmark}
\affiliation[University of Stuttgart (ITT)]
{Institute of Thermodynamics and Thermal Process Engineering (ITT), University of Stuttgart, Pfaffenwaldring 9, 70569 Stuttgart, Germany}
\author{Calin Dan}
\affiliation[University of Stuttgart (ICP)]
{Institute for Computational Physics (ICP), University of Stuttgart, Pfaffenwaldring 27, 70569 Stuttgart, Germany}
\author{Jens Harting}
\affiliation[Eindhoven University of Technology]
{Faculteit Technische Natuurkunde (TN), Eindhoven University of Technology, P.\ O.\ Box 513, 5600 MB Eindhoven, The Netherlands}
\alsoaffiliation[University of Stuttgart (ICP)]
{Institute for Computational Physics (ICP), University of Stuttgart, Pfaffenwaldring 27, 70569 Stuttgart, Germany}
\author{Hans Hasse}
\affiliation[University of Kaiserslautern]
{Laboratory of Engineering Thermodynamics (LTD), University of Kaiserslautern, Erwin-Schr\"odinger-Str.\ 44, 67663 Kaiserslautern, Germany}
\author{Jadran Vrabec}
\email{jadran.vrabec@upb.de}
\affiliation[University of Paderborn]
{Thermodynamics and Energy Technology Laboratory (ThEt), University of Paderborn, Warburger Str.\ 100, 33098 Paderborn, Germany}

\begin{document}

\begin{abstract}
Vapor-liquid menisci of the truncated and shifted Lennard-Jones fluid between parallel planar walls are investigated by molecular dynamics simulation. Thereby, the characteristic energy of the unlike dispersive interaction between fluid molecules and wall atoms is systematically varied to determine its influence on the contact angle. The temperature is varied as well, covering most of the range between the triple point temperature and the critical temperature of the bulk fluid. The transition between obtuse and acute angles is found to occur at a temperature-independent magnitude of the fluid-wall dispersive interaction energy. On the basis of the present simulation results, fluid-wall interaction potentials can be adjusted to contact angle measurements.
\end{abstract}

\section{Introduction}
%
%
A major challenge for molecular modeling consists in the definition of unlike
interaction potentials.
In the past, a variety of combination rules were proposed, none of which was found
to be valid in general. Several of these, including the Lorentz-Berthelot combination
rule, are considered to be a good starting point for further adjustment in most cases \cite{SVH07}.
The present work contributes to understanding the dispersive interaction between a solid wall and
a fluid, which is essential for the analysis of adsorption and microscopic flow properties.

In principle, the Lorentz-Berthelot rule can be applied for effective pair potentials acting
between the fluid particles and the atoms of a solid
wall \cite{Steele78, SPV07}, based on size and energy parameters
derived from properties of the solid and the fluid.
However, while using combination rules to extrapolate from homogeneous bulk
solid and fluid properties to interfacial phenomena can lead to a good agreement
with the actual behavior \cite{Steele78}, this approach has only shaky theoretical
foundations \cite{DLP61}. 
Unlike pair potentials between a fluid and a solid
can only be expected to give
reliable results if they are developed using actual information on
fluid-wall contact effects.
Since adsorption in nanopores can be studied on the basis of
effective pair potentials \cite{Steele74, FF75, FBKF82, SPV07, BYB09, SVH09},
it is obversely possible to fit model parameters 
to adsorption isotherms \cite{VLN99}.
The present study follows the line of research, suggested by \textsc{Werder} \textit{et al.}\ \cite{WWJHK03},
of adjusting unlike parameters to contact angle measurements. 

Using density functional theory, \textsc{Teletzke} \textit{et al.}\ \cite{TSD82} examined
the dependence of wetting and drying transitions on characteristic size
and energy parameters of the fluid-wall dispersive interaction. Subsequently,
\textsc{Soko\l{}owski} and \textsc{Fischer} \cite{SF93} as well as \textsc{Giovambattista} \textit{et al.}\ \cite{GRD06} investigated fluid density profiles in extremely narrow
channels for several values of the fluid-wall dispersive energy
and surface polarity, respectively.
On the microscopic and the nanoscopic level, the statics and dynamics
of fluids under confinement and the corresponding three-phase contact lines can also be investigated 
by the lattice Boltzmann method \cite{HKH06, DBCPS08, DBCPST09, SHHC09}.
%
%

Molecular dynamics (MD) simulation can
be applied to this problem as well, leading to a consistent molecular approach.
The increase in computing power and the development of massively parallel
MD software provide tools for simulating systems with
a large number of particles. System sizes accessible to MD simulation
are getting closer to the smallest experimental settings.
This allows comparing simulation data directly to experimental
observables for a growing spectrum of properties, including the contact angle.
The truncated and shifted Lennard-Jones (\LJTS{})
potential \cite{AT87} is used in the present work for describing both
the fluid-fluid and the fluid-wall interaction, leading to systems that
extend previous studies on interface properties for
the \LJTS{} fluid \cite{VKFH06, HVH08, HVBH09}.
Hysteresis \cite{Monson08} as well as the formation and growth
of liquid precursor layers on the surface \cite{DLP61} are not discussed as
the present work deals with equilibrium properties of the phase boundary only.

\section{Simulation method}

Like the original Lennard-Jones (LJ) potential
$\potentialLJ(\distance) =
   4\LJenergy\left[(\LJsize\slash\distance)^{12}
      - (\LJsize\slash\distance)^{6}\right]$,
the \LJTS{} model \cite{AT87}
\begin{equation}
   \potentialLJTS_{\moli\molj}(\distance_{\moli\molj})
      = \left\{ \begin{array}{l@{\quad \quad}l}
         \potentialLJ(\distance_{\moli\molj})
            - \potentialLJ(\cutoff) & \distance_{\moli\molj} < \cutoff \\
         0 & \distance_{\moli\molj} \geq \cutoff,
      \end{array} \right.
\end{equation}
with a cutoff radius of $\cutoff$ = 2.5 $\LJsize$,
accurately reproduces the thermophysical properties of several non-polar fluids,
in particular noble gases and methane, when adequate values
for the size and energy parameters $\LJsize$ and $\LJenergy$ are speci\-fied \cite{VKFH06}. 
Due to the relatively small cutoff radius, molecular simulation is comparably
fast, while the full descriptive power of the LJ potential is
retained even for systems with phase boundaries \cite{VKFH06}.

%
In order to accurately describe properties of a solid material,
it is usually necessary to employ multi-body potentials which have a large number of model parameters
and are computationally quite expensive \cite{Tersoff88, ESU08, GVLMFF08}.
The present study, however, does not regard the properties of a specific wall material
but rather the influence of the fluid-wall dispersive interaction on the fluid itself.
Accordingly, a layered wall was represented here by a comparably straightforward
system of coupled harmonic oscillators, using different spring
constants $\springconstanty$ and $\springconstantCC$ for
the transverse vibration with respect to the layers
\begin{equation}
   \potentialy_\moli(\ycoord_\moli)
      = \frac{\springconstanty}{2}
         (\ycoord_\moli - \ycoordeq_\moli)^2,
\label{eqn:sol1}
\end{equation}
wherein $\ycoordeq_\moli$ is the equilibrium value of the $\ycoord$
coordinate (i.e.\ the direction perpendicular to the wall), and longitudinal oscillations,
\begin{equation}
   \potentialCC_{\moli\molj}(\distance_{\moli\molj})
      = \frac{\springconstantCC}{2}
         (\distance_{\moli\molj} - \bondlengthCC)^2,
\label{eqn:sol2}
\end{equation}
with respect to the equilibrium bond length $\bondlengthCC$ between
neighboring atoms $\moli$ and $\molj$.

Fluid-wall interactions can be represented by full \cite{Steele74} 
or slightly modified \cite{DYW06} LJ potentials, acting between fluid particles and the atoms of
the solid.
Following this approach, the \LJTS{} potential with the size and
energy parameters $\LJsizeFW = \LJsize$ as well as
\begin{equation}
   \LJenergyFW = \FWenergy\LJenergy,
\end{equation}
was applied for the unlike interaction
using the same cutoff radius as for the fluid.
Potential parameters for the molecular models of the fluid as well as
the solid component were chosen such as to represent methane and
graphite, respectively.
For the fluid, the \LJTS{} size and energy parameters $\LJsize$ = 3.7241 \AA{} and
$\LJenergy\slash\kboltz$ = 175.06 K, as well
as the molecular mass $\molmass$ = 16.04 $\mathrm{u}$
were used \cite{VKFH06}, so that the carbon-carbon (\ce{C}--\ce{C}) bond
length in graphite $\bondlengthCC$ = 1.421 \AA{} \cite{HW01} can
be expressed in LJ units as $\bondlengthCC$ = 0.3816 $\LJsize$, while
the interlayer distance $\interlayer$ = 3.35 \AA{} corresponds to
0.8996 $\LJsize$.

The spring constant $\springconstantCC$ = 15600 N/m related to the \sptwo{} bonds
was adjusted to the \ce{C}--\ce{C} radial distribution function obtained from simulations with
the \textsc{Tersoff} \cite{Tersoff88} potential. This distribution had to be rescaled
because as previously shown \cite{Kelires93, HVBH09}, the Tersoff potential deviates 
by about 3\%{} from the actual bond length in graphite.
In agreement with the relation between the \ce{C}--\ce{C} bond energy (4.3 eV)
and the interaction energy between adjacent graphite layers (0.07 eV) \cite{CWG94},
the interlayer spring constant was specified
as $\springconstanty = \springconstantCC\slash{}60$.
Equations (\ref{eqn:sol1}) and (\ref{eqn:sol2}) ensure
that the wall atoms oscillate around the fixed $\ycoord$ coordinate that corresponds
to their layer, while no particular $\xcoord$ and $\zcoord$ coordinates are preferred
because the atoms are only connected with their immediate neighbors to permit
individual sliding of the wall layers.

\begin{figure}[t!]
\includegraphics[width=8.25cm]{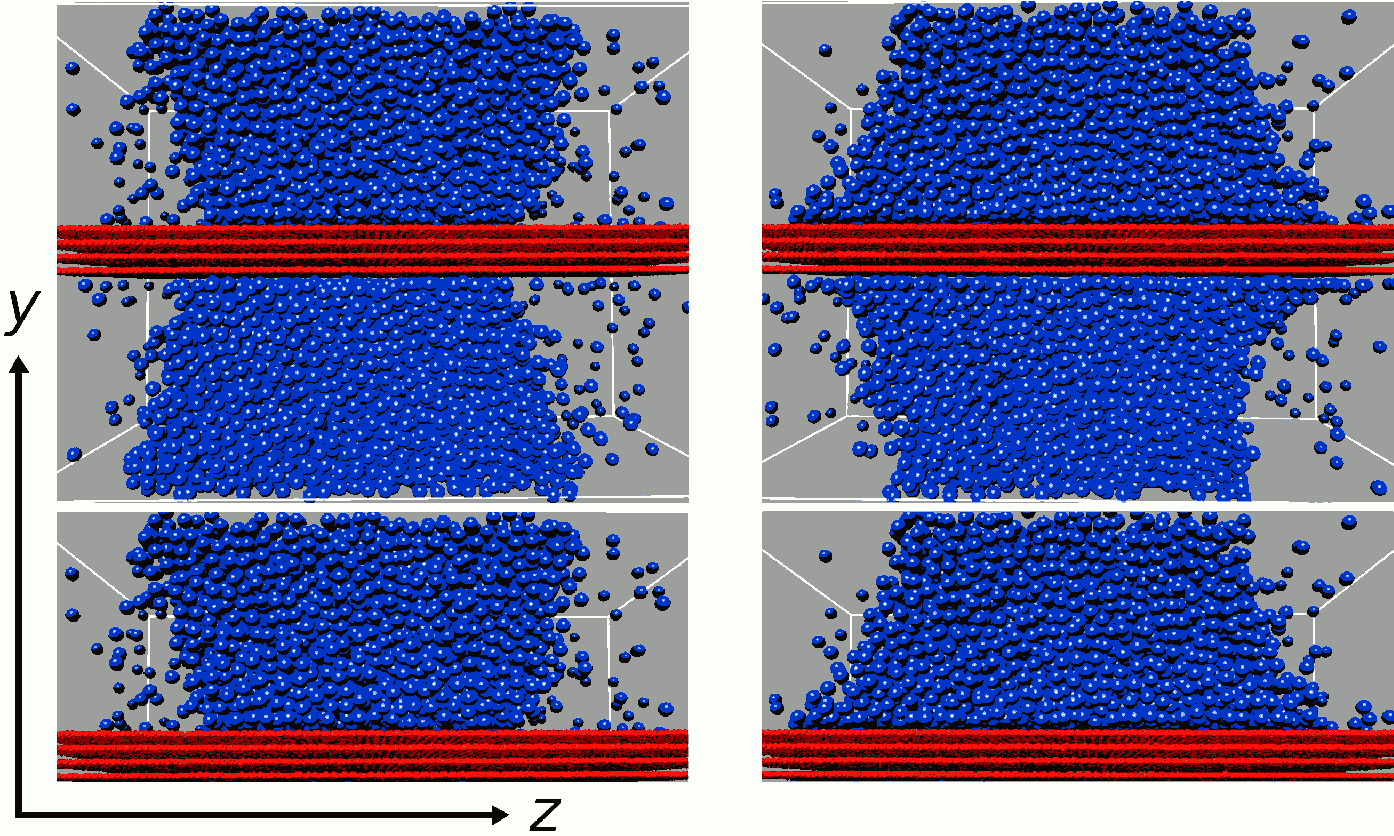}
\caption{Simulation snapshots for the reduced fluid-wall dispersive energy
$\FWenergy$ = 0.09 (left) and 0.16 (right) at a temperature $\temperature$ = 0.73 $\LJenergy\slash\kboltz$.
The upper half is reproduced in the bottom to illustrate the effect of the periodic
boundary condition.}
\label{fig04}
\end{figure}

Massively parallel canonical ensemble MD simulations were conducted with the
program \mardyn{} \cite{BNHVDHB09} to obtain the contact angle dependence on
the temperature and the reduced fluid-wall dispersive energy.
For all simulation runs, the integration time step was specified as 1 fs. \Vapour{} and
liquid were independently equilibrated in homogeneous simulations for 10 ps.
This was followed by
200 ps of equilibration for the combined system, i.e.\ a liquid meniscus surrounded
by \vapor{}, with a wall consisting of four to seven
layers, cf.\ \ref{fig04}, where the starting configuration contained a planar
\vapor-liquid interface perpendicular to the $\zcoord$ coordinate axis. Note that
the distance from the wall is given by the $\ycoord$ coordinate, while $\zcoord$
is the characteristic direction for the density gradient of the fluid.
The periodic boundary condition was applied to the system, leaving a channel for the
fluid with a height of 27 $\LJsize$ between the wall and its periodic image.

\begin{figure}[h!]
\includegraphics[width=8.25cm]{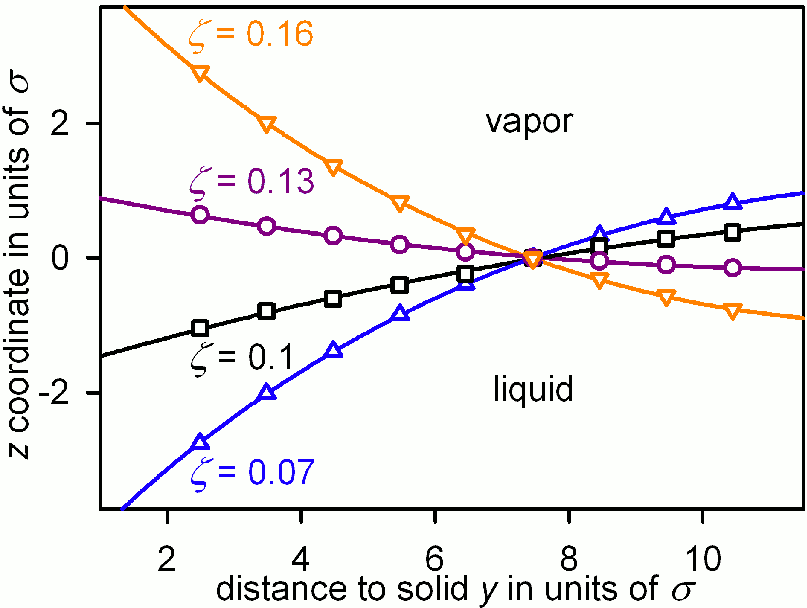}
\caption{\Vapour{}-liquid interface profiles for the reduced fluid-wall dispersive
energy $\FWenergy$ = 0.07 (upward triangles), 0.10 (squares), 0.13 (circles), and
0.16 (downward triangles) at a temperature $\temperature$ = 0.82 $\LJenergy\slash\kboltz$.
Note that the full lines are circle segments, adjusted to the data points that
are represented by symbols. The almost perfect match between the individual simulation
results, indicated by their collective agreement with the fit, reflects the precision
of the present simulation data. At $\temperature$ = 1 $\LJenergy\slash\kboltz$, however, the
margin of error becomes more significant.}
\label{fig0608}
\label{fig06}
\label{fig08}
\end{figure}

Via binning, the density profiles were averaged
over at least 800 ps after equilibration. The arithmetic mean density
$\densitytr = (\densityliq + \densityvap) / 2$
was applied to define the position of the phase boundary, where
$\densityliq$ and $\densityvap$ are the saturated bulk densities of liquid
and \vapour{} which are known for the \LJTS{} fluid from previous work \cite{VKFH06}.
In the immediate vicinity of the wall, the fluid
is affected by short-range ordering effects \cite{KS91, HKH06, BYB09}.
The influence of this phenomenon was minimized by taking
density averages over a bin size of about 1 $\LJsize$, cf.\ \ref{fig08},
following \textsc{Giovambattista} \textit{et al.}\ \cite{GDR07}. 
A circle was adjusted to the positions of the interface in the bins
corresponding to distances between 3 and 11 $\LJsize$ from the wall,
and the tangent to this circle at a distance of 1 $\LJsize$ from
the wall was consistently used to determine the contact angle.

\section{Simulation results}

Menisci between parallel planar walls were simulated 
for a reduced fluid-wall dispersive energy $\FWenergy$ between 0.07 and 0.16
at temperatures of 0.73, 0.82, 0.88, and 1 $\LJenergy\slash\kboltz$.
Note that for the bulk \LJTS{} fluid, the triple point temperature
is about 0.65 $\LJenergy\slash\kboltz$ according to \textsc{van Meel} \textit{et al.}\ \cite{MPSF08}
while the critical temperature is 1.0779 $\LJenergy\slash\kboltz$ according to \textsc{Vrabec} \textit{et al.}\ \cite{VKFH06},
so that almost the entire regime of \vapour-liquid coexistence was
covered here.

High values of $\FWenergy$ correspond to a strong attraction between fluid particles
and wall atoms, leading to a contact angle $\contactangle < \degree{90}$, 
i.e.\ to partial ($\contactangle > \degree{0}$) or perfect ($\contactangle = \degree{0}$)
wetting of the surface.
As expected, with increasing fluid-wall dispersive energy, the extent of
wetting grows, cf.\ \ref{fig04}.
As can be seen in \ref{tab6} and \ref{fig0608}, the transition
from obtuse to acute contact angles
occurs at $\FWenergy$ values between 0.11 and 0.13 over the whole studied
temperature range. Present simulation results were correlated by
\begin{eqnarray}
   \cos\contactangle(\temperature, \FWenergy)
      &=& \left(1 + \frac{\inftemperature^{1.7}}{27}\right)
         \tanh \frac{\FWenergy - \cpenergy}{0.087},
\label{eqn:corr1}
\end{eqnarray}
where $\inftemperature = (1 - \temperature\slash\temperaturecrit)^{-1}$
approaches infinity for $\temperature \to \temperaturecrit$, while
$\cpenergy = 0.119$ is the reduced fluid-wall dispersive energy
that leads to a contact angle of $\contactangle = \degree{90}$.

\begin{table}[b!]
\caption{ Contact angle of the \LJTS{} fluid on graphite from MD simulation
          as a function of reduced fluid-wall dispersive energy and temperature. }
\label{tab6}
\begin{center}
\begin{tabular}{l|cccc}
 \backslashbox{$\FWenergy$}{$\kboltz\temperature\slash\LJenergy$} & 0.73 & 0.82 & 0.88 & 1 \\ \hline
 0.07 & $\degree{127}$ & $\degree{134}$
      & $\degree{139}$ & $\degree{180}$ \\
 0.09 & $\degree{112}$ & $\degree{116}$
      & $\degree{119}$ & $\degree{180}$ \\
 0.10 & $\degree{107}$ & $\degree{106}$
      & $\degree{109}$ & $\degree{145}$ \\
 0.11 & $\degree{\phantom{0}99}$ & $\degree{\phantom{0}95}$
      & $\degree{\phantom{0}96}$ & $\degree{128}$ \\
 0.12 & &
      & $\degree{\phantom{0}86}$ & $\degree{\phantom{0}86}$ \\
 0.13 & $\degree{\phantom{0}82}$ & $\degree{\phantom{0}79}$
      & $\degree{\phantom{0}76}$ & $\degree{\phantom{0}81}$ \\
 0.14 & $\degree{\phantom{0}73}$ & $\degree{\phantom{0}67}$
      & $\degree{\phantom{0}63}$ & $\degree{\phantom{00}0}$ \\
 0.16 & $\degree{\phantom{0}58}$ & $\degree{\phantom{0}45}$
      & $\degree{\phantom{0}40}$ & $\degree{\phantom{00}0}$ \\
 \hline
\end{tabular}
\end{center}
\end{table}

\begin{figure}[h!]
\includegraphics[width=8.25cm]{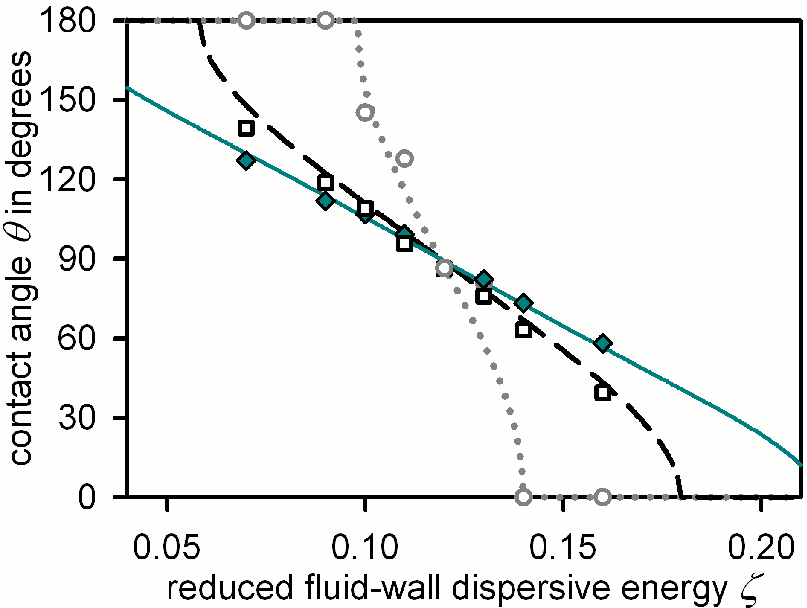}
\caption{Simulation results and correlation, cf.\ Eq.\ (\ref{eqn:corr1}),
for the contact angle
in dependence of the reduced fluid-wall dispersive energy $\FWenergy$ 
at temperatures $\temperature$ = 0.73 (diamonds and solid line), 0.88 (squares and dashed line)
as well as 1 $\LJenergy\slash\kboltz$ (circles and dotted line).}
\label{fig13}
\end{figure}

Perfect wetting or drying is present where
Eq.\ (\ref{eqn:corr1}) yields $\cos\contactangle \geq 1$
or $\cos\contactangle \leq -1$, respectively.
In particular, both the value of $\cpenergy$ and the symmetry relation
\begin{equation}
   \cos\contactangle(\temperature, \cpenergy - \Delta\FWenergy)
      = - \cos\contactangle(\temperature, \cpenergy + \Delta\FWenergy),
\label{eqn:sym}
\end{equation}
were found to be temperature-independent.
Note that the simulation results stronlgy suggest such a symmetry property,
cf.\ \ref{fig13}, which is not an artifact of the correlation.

\begin{figure}[t!]
\includegraphics[width=8.25cm]{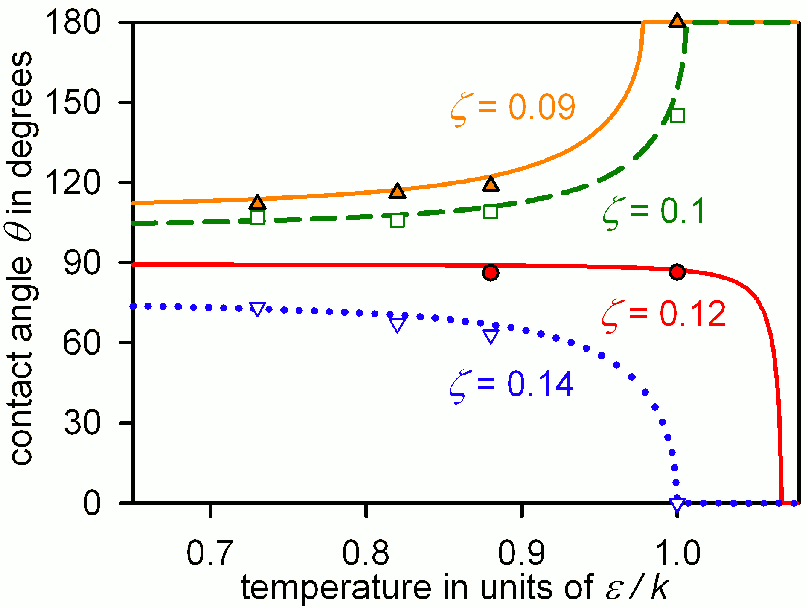}
\caption{Simulation results and correlation, cf.\ Eq.\ (\ref{eqn:corr1}),
for the contact angle
in dependence of the temperature at fluid-wall dispersive energies $\FWenergy$ 
= 0.09 (upward triangles and solid line), 0.10 (squares and dashed line),
0.12 (bullets and solid line) as well as 0.14 (downward triangles and dotted line).
The entire temperature range between triple point and critical point
of the bulk fluid is shown.}
\label{fig11}
\end{figure}

\ref{fig13} shows, particularly for high temperatures, that there is a narrow
range of $\FWenergy$ values that lead to the
formation of a contact angle as opposed to perfect wetting or drying.
The present plots agree quali\-tatively
with those determined by \textsc{Giovambattista} \textit{et al.}\ \cite{GDR07} for the influence
of the polarity of hydroxylated silica surfaces on the contact angle formed
with water. In \ref{fig11}, it can be seen that the extent of
wetting (for $\FWenergy > \cpenergy$) or drying (for
$\FWenergy < \cpenergy$), respectively, increases as the temperature approaches
$\temperaturecrit$. Eventually, this leads to the known
phenomenon of critical point wetting\cite{Cahn77} for the whole range
above a wetting temperature $\temperaturew$.
The effects described above can be accounted for by
the Young equation
\begin{equation}
   \cos\contactangle(\temperature, \FWenergy) =
      \frac{\tensions(\temperature, \FWenergy, \densityvap)
         - \tensions(\temperature, \FWenergy, \densityliq)}{\tensionvl(\temperature)},
\end{equation}
which relates the \vapour-liquid surface tension $\tensionvl$ and
the contact angle to the interfacial tension $\tensions$ that acts
between wall and \vapour{} or liquid. 
The deviation from $\contactangle = \degree{90}$
increases with $\temperature \to \temperaturecrit$
because $\tensionvl(\temperature)$ converges to zero faster than
the density difference $\densityliq(\temperature) - \densityvap(\temperature)$
between the two fluid phases.
The relevant critical exponents in case of the LJTS fluid are
\begin{equation}
   \lim_{\temperature\to\temperaturecrit}
      \frac{\ln\left[\densityliq(\temperature) - \densityvap(\temperature)\right]}{
         \ln(\temperaturecrit - \temperature)} \approx \frac{1}{3},
\end{equation}
for the saturated densities of the bulk fluid \cite{VKFH06}, in accordance with the
\textsc{Guggenheim} \cite{Guggenheim45} approach, and
\begin{equation}
      \frac{\differential \ln\tensionvl(\temperature)}{
         \differential \ln(\temperaturecrit - \temperature)} \approx 1.21,
\end{equation}
for the \vapour{}-liquid surface tension \cite{VKFH06}, confirming a
similar value (1.26) obtained from fluctuation theory of critical
phenomena \cite{Anisimov87}.

The correlation given by Eq.\ (\ref{eqn:corr1})
suggests for the present system a first-order transition between partial and perfect wetting
or drying, respectively, as described by
\textsc{Cahn} \cite{Cahn77}.
With $\FWenergy \approx \cpenergy$, \ref{fig13} shows that the contact angle
$\contactangle$ depends linearly on the fluid-wall dispersive energy, and the symmetry
property suggests for $\FWenergy = \cpenergy$ that $\tensions$ does
not depend on the density of the fluid.

\section{Comparative discussion}

Qualitatively, the symmetry relation given by Eq.\ (\ref{eqn:sym})
corroborates \textsc{Monson} \cite{Monson08} who obtained the same property
based on mean-field DFT calculations.
If each layer of the wall is approximated as a plane of
uniform density $\surfacedensity$, the well depth of the fluid-wall dispersive
interaction, which can be used to compare different interaction models quantitatively, is given by
\begin{equation}
   \welldepth = - \surfacedensity \FWenergy \,\, \min_{\ycoord > 0}
      \int_0^\infty \differential\surfacerad \,\, 2\pi\surfacerad
         \sum_{\layer = 0}^{\layers-1}
            \potentialLJTS\left(\left[(\ycoord + \layer\interlayer)^2
               + \surfacerad^2\right]^{1\slash{}2}\right),
\label{eqn:welldepth}
\end{equation}
for a system of $\layers$ layers with an interlayer distance
of $\interlayer$. In the present case with $\cutoff$ = 2.5 $\LJsize$
for the \LJTS{} potential as well as a surface density of $\surfacedensity$ $=$ 5.287 $\LJsize^{-2}$,
an interlayer distance of $\interlayer$ = 0.8996 $\LJsize$, and the number of layers
$\layers$ = 3 (the fourth layer of the wall is beyond the cutoff radius), one obtains
\begin{equation}
   \welldepth = 17.29 \, \kboltz\temperaturecrit \,\mult\, \FWenergy,
\label{eqn:presentwelldepth}
\end{equation}
normalized by the critical temperature $\temperaturecrit$ =
1.0779 $\LJenergy\slash\kboltz$ of the bulk \LJTS{} fluid \cite{VKFH06}.
The transition from obtuse to acute contact angles, occurring at $\cpenergy = 0.119$
in the present case, therefore corresponds to a well depth
of $\welldepth_\cpenergy$ = 2.057 $\kboltz\temperaturecrit$.

A very similar system was investigated by \textsc{Bucior} \textit{et al.}\ \cite{BYB09}, albeit
for the \LJTS{} fluid with a cutoff radius of $2^{7\slash{}6}$ $\LJsize$
as well as a rigid wall with a surface density
$\surfacedensity$ = 1.7342 $\LJsize^{-2}$ and a single solid layer, i.e.\ $\layers$ $=$ 1.
For that fluid, the critical temperature is given by \textsc{Schrader} \textit{et al.}\ \cite{SVB09} as 
$\temperaturecrit$ $=$ $0.9999$ $\LJenergy\slash\kboltz$,
so that the fluid-wall dispersive energy was related to the well
depth by $\welldepth$ = 5.429 $\kboltz\temperaturecrit$ $\mult$ $\FWenergy$
according to Eq.\ (\ref{eqn:welldepth}). From the density profiles
of \textsc{Bucior} \textit{et al.}\ \cite{BYB09} one finds that a rectangular contact angle is
reached for a reduced fluid-wall dispersive energy $\FWenergy$ between 0.61 and 0.7, i.e.\
$3.3 \, \kboltz\temperaturecrit < \welldepth_\cpenergy < 3.8 \, \kboltz\temperaturecrit$
which is on the same order of magnitude as the present result. The
quantitative deviation has to be attributed to the different solid structure,
since the single wall layer of \textsc{Bucior} \textit{et al.}\ \cite{BYB09} leads to a faster decay
of the fluid-wall dispersion with respect to the distance than in
the present simulations where three layers can directly interact with
the fluid. A larger value of $\welldepth_\cpenergy \slash (\kboltz\temperaturecrit)$
is required to compensate for the effectively smaller length scale
of the dispersive interaction.

The present correlation, cf.\ Eq.\ (\ref{eqn:corr1}), predicts perfect wetting for
\begin{equation}
   \temperaturew\left(\frac{\welldepth}{\kboltz\temperaturecrit}\right)
      = \left(1 - 0.144 \left[\left(\tanh
         \left[0.665 \frac{\welldepth}{\kboltz\temperaturecrit} - 1.37\right]\right)^{-1} - 1\right]^{-0.588}\right) \temperaturecrit.
\end{equation}
The transition to perfect wetting was also simulated by \textsc{Bojan} \textit{et al.}\ \cite{BSCSC99}
who applied the Monte Carlo method in the grand canonical ensemble
to neon on metal surfaces. Thereby, the full LJ fluid,
with $\temperaturecrit$ = 1.310 $\LJenergy\slash\kboltz$ as determined by \textsc{Lotfi} \textit{et al.}\ \cite{LVF92},
was used to model neon.
With $\welldepth$ = 2.13 $\kboltz\temperaturecrit$, representing
magnesium, they obtained
a wetting temperature of $\temperaturew$ $\approx$ 0.50 $\temperaturecrit$,
as opposed to the present results which imply that
perfect wetting is only reached above 0.974 $\temperaturecrit$,
in the immediate vicinity of the critical temperature.

As \textsc{Bojan} \textit{et al.}\ \cite{BSCSC99} themselves remark, their calculations predict a much
lower wetting temperature than a similar previous study by \textsc{Soko\l{}owski} and \textsc{Fischer} \cite{SF90}
on the local structure of fluid argon in contact with solid carbon
dioxide (with the same value of $\welldepth\slash\LJenergy$). The latter MD simulation results
used a size parameter $\LJsizeFW\slash\LJsize$ that is significantly smaller than unity
in case of argon and carbon dioxide \cite{SF90, KS91}
and therefore cannot be directly compared to the present work, since varying the interaction
length scale can lead to qualitatively different properties such as a change
in the order of the wetting transition \cite{TSD82}.
However, it should be pointed out that significantly better agreement was obtained
here with the results of \textsc{Soko\l{}owski} and \textsc{Fischer} \cite{SF90} than with those of \textsc{Bojan} \textit{et al.}\ \cite{BSCSC99}.

The results presented above can now be used to provide an estimate for
the magnitude of the dispersive interaction between fluids
and wall materials for which experimental data on the contact angle are
available. For instance, regarding the refrigerant R134a (1,1,1,2-tetrafluoroethane)
at temperatures between 10 and 80 $^\circ$C,
\textsc{Vadgama} and \textsc{Harris} \cite{VH07} obtained contact angles of 5.5$^\circ$ $\pm$ 1$^\circ$ on
copper and 7$^\circ$ $\pm$ 1$^\circ$ on aluminum. With
$\temperature$ = 0.85 $\temperaturecrit$, which is 45 $^\circ$C for R134a,
the well depth can be estimated as
$\welldepth\slash\kboltz$ $\approx$ 2.9 $\temperaturecrit$ = 1100 K
in both cases
on the basis of Eqs.\ (\ref{eqn:corr1}) and (\ref{eqn:presentwelldepth}).

This can be related to a molecular model of the dispersive interaction by
\begin{equation}
   \welldepth = - \soldensity \,\, \min_{\ycoord > 0}
      \int_0^\infty \differential\surfacerad \,\, 2\pi\surfacerad
         \int_{0}^{\infty} \differential\ycoordadd \,\,
            \potentialfw\left(\left[(\ycoord + \ycoordadd)^2
               + \surfacerad^2\right]^{1\slash{}2}\right),
\label{eqn:LJ-9-3}
\end{equation}
wherein $\soldensity$ is the density of the solid wall and
$\potentialfw$ is the dispersive interaction potential
acting between a fluid molecule and a wall atom.
Note that, while Eq.\ (\ref{eqn:welldepth}) corresponds to a sum over truncated and shifted LJ-10-4 terms,
Eq.\ (\ref{eqn:LJ-9-3}) does not rely on any particular
assumption on the internal structure of the solid wall. If e.g.\ a LJ-12-6 potential is used
for $\potentialfw$, it corresponds to a LJ-9-3 interaction.
This reasoning can plausibly be applied to all fluids that do
not exhibit an excessively polar or anisotropic structure.

\section{Conclusion}
The contact angle formed between a wall and a \vapour-liquid interface
was determined by canonical ensemble MD simulation,
where the magnitude of the dispersive fluid-wall interaction and the temperature were varied. Over the whole
temperature range under investigation,
the contact angle dependence on the fluid-wall dispersive energy was
found to follow a simple symmetry law.
At a temperature-independent value of the reduced fluid-wall dispersive energy,
the interfacial tension between \vapour{} and solid as well as liquid and solid is equal,
corresponding to the transition between obtuse and acute contact angles.

\begin{acknowledgement}
The authors thank
Deutsche Forschungsgemeinschaft for funding SFB 716,
the German Federal Ministry of Education and Research (BMBF)
for funding the IMEMO project, and
the D-Grid Initiative as well as the High Performance Computing Center Stuttgart (HLRS)
for providing access to the \textit{gt4bw} supercomputer (D-Grid resource \textit{hlrs-bwgrid}).
Furthermore, they thank G.\ K.\ Auernhammer and M.\ Sokuler (MPI f\"ur Polymerforschung),
F.\ G\"ahler (Universit\"at Bielefeld), M.\ Hecht (HLRS) as well as G.\ C.\ Lehmann (Universit\"at Paderborn) for discussions
and M.\ Buchholz (TU M\"unchen) as well as M.\ Bernreuther, D.\ Jenz, and
C.\ Niethammer (HLRS) for their contribution to de\-veloping the
\mardyn{} program. The presented research
was conducted under the auspices of the Boltzmann-Zuse Socie\-ty of
Computational Molecular Engineering (BZS).
\end{acknowledgement}

\bibliography{LJTScontactangle2010}


\end{document}